\documentclass{Interspeech}

\usepackage{url}
\usepackage{arydshln}
\usepackage{svg}
\usepackage{multirow}

\interspeechcameraready

\title{Towards Pre-training an Effective Respiratory Audio Foundation Model}

\author[affiliation={1}]{Daisuke}{Niizumi}
\author[affiliation={1}]{Daiki}{Takeuchi}
\author[affiliation={1}]{Masahiro}{Yasuda}
\author[affiliation={1}]{Binh Thien}{Nguyen}
\author[affiliation={1}]{Yasunori}{Ohishi}
\author[affiliation={1}]{Noboru}{Harada}

\affiliation{NTT Communication Science Laboratories}{NTT Corporation}{Japan}
\email{daisuke.niizumi@ntt.com, d.takeuchi@ntt.com, masahiro.yasuda@ntt.com, binhthien.nguyen@ntt.com, yasunori.ohishi@ntt.com, harada.noboru@ntt.com}
\keywords{audio foundation model, pre-training, respiratory sound}

\usepackage{comment}

\begin{document}

\maketitle

\begin{abstract}
Recent advancements in foundation models have sparked interest in respiratory audio foundation models. However, the effectiveness of applying conventional pre-training schemes to datasets that are small-sized and lack diversity has not been sufficiently verified. This study aims to explore better pre-training practices for respiratory sounds by comparing numerous pre-trained audio models.
Our investigation reveals that models pre-trained on AudioSet, a general audio dataset, are more effective than the models specifically pre-trained on respiratory sounds. Moreover, combining AudioSet and respiratory sound datasets for further pre-training enhances performance, and preserving the frequency-wise information when aggregating features is vital. Along with more insights found in the experiments, we establish a new state-of-the-art for the OPERA benchmark, contributing to advancing respiratory audio foundation models. Our code is available online.\footnote{\url{https://github.com/nttcslab/eval-audio-repr/tree/main/plugin/OPERA}} 
\end{abstract}

\section{Introduction}
As a trend in non-invasive diagnostic methods using sound~\cite{Cook2022Body}, respiratory audio foundation models have gained attention, driven by recent advancements in deep learning foundation models~\cite{baur2024HeAR,FM_cardiovascular,OPERA}. These models may serve as a fundamental block for health monitoring and disease diagnosis applications. Since training foundation models requires large-scale data, existing methods have curated their respiratory sound datasets, trained models using established pre-training schemes, and demonstrated their effectiveness through benchmarks~\cite{OPERA,AuscultaBase}.

However, the datasets used for training these respiratory audio models are small and lack diversity compared to those for general audio models.
While general audio datasets contain a wide variety of sounds on a large scale, respiratory audio datasets are more than ten times smaller than general audio ones~\cite{OPERA,AuscultaBase}, and its sounds, such as coughs and breaths, are monotonic.
In addition, the effectiveness of the conventional pre-training schemes using such a small and monotonic dataset is unclear.

Towards pre-training a respiratory audio foundation model that is practically effective, we empirically investigate what makes the pre-trained models advantageous for respiratory sound analysis.
To do this, we evaluate numerous pre-trained audio models through a unified respiratory task benchmark and compare the results to explore better practices. The various models have taken different pre-training approaches; thus, comparing them on a unified benchmark can surface the factors of effective pre-training for respiratory sounds.
Our research questions are the following.

\begin{itemize}
\item \textbf{RQ1} What are the better practices for pre-training effective respiratory audio foundation models?
\item \textbf{RQ2} What data lead to learning effective features?
\item \textbf{RQ3} Are the intermediate layer features effective?
\end{itemize}

The experiments reveal various insights, such as the advantage of combining a diverse large-scale dataset with respiratory datasets and the importance of preserving frequency-wise information in feature aggregation.
In addition, our results renew the state-of-the-art (SOTA) performance on the OPERA benchmark with a large margin. Our evaluation code is available online for future advancement of respiratory audio foundation models.

\section{Experimental Setup} \label{sec:exp-setup}
We utilized a unified respiratory audio benchmark to evaluate models under the following experimental setup.

\subsection{Benchmark} \label{sec:exp-benchmark}

\begin{table}[t]
\vspace{-5pt}
\centering
\caption{OPERA benchmark tasks used in our study.}
\label{tab:opera_benchmark}
\vspace{-5pt}
\resizebox{1.0\columnwidth}{!}{%
\begin{tabular}{rllll}
\toprule
Dataset & ID & Task & Modality & Class samples \\
\midrule
COUGHVID~\cite{COUGHVID} & T5 & Covid/Non-covid & Cough & 547 / 5628 \\
         & T6 & Female/Male & Cough & 2468 / 4795 \\
ICBHI~\cite{ICBHI2017}    & T7 & COPD$^\dagger$/Healthy & Lung sounds & 793 / 35 \\
Coswara~\cite{Coswara}  & T8 & Smoker/Non-smoker & Cough & 201 / 747 \\
         & T9 & Female/Male & Cough & 759 / 1737 \\
KAUH~\cite{KAUH}     & T10& Obstructive/Healthy & Lung sounds & 129 / 105 \\
Resp.@TR~\cite{Resp_at_tr} & T11& COPD$^\dagger$ severity (0-4) & Lung sounds & 72/60/84/84/204 \\
\bottomrule
\addlinespace[0.05cm]
\multicolumn{5}{l}{$^{\dagger}$Chronic obstructive pulmonary disease, a common lung disease.}\\
\end{tabular}
}
\vspace{-10pt}
\end{table}

We conducted experiments on the publicly available OPERA \cite{OPERA} benchmark and used seven tasks (T5 to T11) for which the data are publicly available. The seven tasks listed in Table \ref{tab:opera_benchmark} are health condition classification problems, where all are binary classification tasks except T11, which is a five-class classification. The benchmark takes a linear evaluation protocol, where model weights are frozen. A model encodes task data samples into features, and a linear layer is trained to classify the task using the features as input. The performance metric is AUROC (Area Under the Receiver Operating Characteristic), and the results are the statistics for five attempts.

\begin{table*}[tb!]
\vspace{-10pt}
\caption{Performance comparison among audio foundation models on OPERA benchmark.}
\label{tab:exp:result-opera}
\vspace{-5pt}
\centering
\resizebox{0.9\textwidth}{!}{%
\begin{tabular}{lllllllll}
\toprule
\multicolumn{1}{r}{Task} & T5 & T6 & T7 & T8 & T9 & T10 & T11 & \\
Model & Covid & Gender & COPD & Smoker & Gender & Obstructive & COPD (5 cls) & Avg.\\
\midrule
\multicolumn{8}{l}{\textit{Supervised learning models.}}  \\
1. AST~\cite{gong2021ast} & 0.607 {\scriptsize $\pm$0.008}$^*$ & 0.739 {\scriptsize $\pm$0.002} &\textbf{1.000 {\scriptsize $\pm$0.000}}$^*$& 0.671 {\scriptsize $\pm$0.011} & 0.833 {\scriptsize $\pm$0.000} & 0.837 {\scriptsize $\pm$0.007}$^*$ & 0.652 {\scriptsize $\pm$0.020}$^*$ & 0.763$^*$\\
2. PANNs Cnn14~\cite{kong2020panns} & 0.533 {\scriptsize $\pm$0.006} & 0.566 {\scriptsize $\pm$0.003} & 0.606 {\scriptsize $\pm$0.037} & 0.533 {\scriptsize $\pm$0.014} & 0.549 {\scriptsize $\pm$0.007} & 0.447 {\scriptsize $\pm$0.057} & 0.527 {\scriptsize $\pm$0.022} & 0.537\\
3. HTS-AT~\cite{Chen2022HTS-AT} & 0.577 {\scriptsize $\pm$0.002} & 0.615 {\scriptsize $\pm$0.001} & 0.765 {\scriptsize $\pm$0.013} & 0.590 {\scriptsize $\pm$0.010} & 0.699 {\scriptsize $\pm$0.002} & 0.778 {\scriptsize $\pm$0.014}$^*$ & 0.521 {\scriptsize $\pm$0.031} & 0.649\\
\multicolumn{8}{l}{\textit{Speech SSL models (with the best performing layer used for extracting features).}}  \\
4. wav2vec2$_\text{ Layer\#7}$~\cite{baevski2020wav2vec2} & 0.480 {\scriptsize $\pm$0.005} & 0.634 {\scriptsize $\pm$0.003} & 0.172 {\scriptsize $\pm$0.013} & 0.589 {\scriptsize $\pm$0.020} & 0.606 {\scriptsize $\pm$0.004} & 0.620 {\scriptsize $\pm$0.021} & 0.560 {\scriptsize $\pm$0.022} & 0.523\\
5. HuBERT$_\text{ Layer\#7}$~\cite{Hsu2021HuBERT} & 0.558 {\scriptsize $\pm$0.002} & 0.736 {\scriptsize $\pm$0.001} & 0.644 {\scriptsize $\pm$0.012} & 0.683 {\scriptsize $\pm$0.004} & 0.807 {\scriptsize $\pm$0.002} & 0.689 {\scriptsize $\pm$0.019} & 0.658 {\scriptsize $\pm$0.018}$^*$ & 0.682\\
6. WavLM$_\text{ Layer\#6}$~\cite{Chen2022WavLM} & 0.555 {\scriptsize $\pm$0.002} & 0.700 {\scriptsize $\pm$0.001} & 0.599 {\scriptsize $\pm$0.016} & 0.687 {\scriptsize $\pm$0.004}$^*$ & 0.771 {\scriptsize $\pm$0.001} & 0.703 {\scriptsize $\pm$0.020} & 0.624 {\scriptsize $\pm$0.011} & 0.663\\
\multicolumn{8}{l}{\textit{CLAP models.}}  \\
7. LAION-CLAP~\cite{LAION-CLAP} & 0.549 {\scriptsize $\pm$0.001} & 0.660 {\scriptsize $\pm$0.001} & 0.674 {\scriptsize $\pm$0.062} & 0.531 {\scriptsize $\pm$0.003} & 0.714 {\scriptsize $\pm$0.002} & 0.776 {\scriptsize $\pm$0.015}$^*$ & 0.584 {\scriptsize $\pm$0.033} & 0.641\\
8. CLAP$_{2022}$~\cite{CLAP2022} & 0.599 {\scriptsize $\pm$0.007}$^*$ & 0.665 {\scriptsize $\pm$0.001} & 0.933 {\scriptsize $\pm$0.005}$^*$ & 0.680 {\scriptsize $\pm$0.009} & 0.742 {\scriptsize $\pm$0.001} & 0.697 {\scriptsize $\pm$0.004} & 0.636 {\scriptsize $\pm$0.045}$^*$ & 0.707\\
9. CLAP$_{2023}$~\cite{CLAP2023} & 0.602 {\scriptsize $\pm$0.007}$^*$ & 0.779 {\scriptsize $\pm$0.001} & 0.988 {\scriptsize $\pm$0.004}$^*$ & 0.687 {\scriptsize $\pm$0.007}$^*$ & 0.866 {\scriptsize $\pm$0.001} & 0.795 {\scriptsize $\pm$0.012}$^*$ & 0.606 {\scriptsize $\pm$0.037} & 0.760$^*$\\
\multicolumn{8}{l}{\textit{General audio SSL models.}}  \\
10. BYOL-A~\cite{niizumi2023byol-a} & 0.531 {\scriptsize $\pm$0.011} & 0.702 {\scriptsize $\pm$0.006} & 0.950 {\scriptsize $\pm$0.031}$^*$ & 0.581 {\scriptsize $\pm$0.035} & 0.807 {\scriptsize $\pm$0.011} & 0.710 {\scriptsize $\pm$0.047} & 0.566 {\scriptsize $\pm$0.029} & 0.693\\
11. ATST-Clip~\cite{Li2023ATST-TALSP} & 0.609 {\scriptsize $\pm$0.009}$^*$ & 0.793 {\scriptsize $\pm$0.001} & 0.977 {\scriptsize $\pm$0.007} & 0.679 {\scriptsize $\pm$0.010} & 0.850 {\scriptsize $\pm$0.002} & 0.828 {\scriptsize $\pm$0.012} & 0.606 {\scriptsize $\pm$0.029} & 0.763$^*$\\
12. ATST-Frame~\cite{Li2023ATST-TALSP} &\textbf{0.621 {\scriptsize $\pm$0.007}}$^*$& \textbf{0.801 {\scriptsize $\pm$0.001}}$^*$ & 0.998 {\scriptsize $\pm$0.001}$^*$ & 0.687 {\scriptsize $\pm$0.010}$^*$ & \textbf{0.908 {\scriptsize $\pm$0.001}}$^*$ &\textbf{0.843 {\scriptsize $\pm$0.006}}$^*$& 0.657 {\scriptsize $\pm$0.003}$^*$ & \textbf{0.788}$^*$\\
13. AudioMAE~\cite{huang2022audiomae} & 0.554 {\scriptsize $\pm$0.004} & 0.628 {\scriptsize $\pm$0.001} & 0.886 {\scriptsize $\pm$0.017}$^*$ & 0.549 {\scriptsize $\pm$0.022} & 0.724 {\scriptsize $\pm$0.001} & 0.616 {\scriptsize $\pm$0.041} & 0.510 {\scriptsize $\pm$0.021} & 0.638\\
14. BEATs~\cite{chen2022beats} & 0.555 {\scriptsize $\pm$0.002} & 0.644 {\scriptsize $\pm$0.001} & 0.823 {\scriptsize $\pm$0.011} & 0.631 {\scriptsize $\pm$0.004} & 0.695 {\scriptsize $\pm$0.003} & 0.723 {\scriptsize $\pm$0.031}$^*$ & 0.623 {\scriptsize $\pm$0.015} & 0.670\\
15. MSM-MAE~\cite{niizumi2022msm-mae} & 0.569 {\scriptsize $\pm$0.003} & 0.781 {\scriptsize $\pm$0.000} &\textbf{1.000 {\scriptsize $\pm$0.000}}$^*$& \textbf{0.721 {\scriptsize $\pm$0.008}}$^*$ & 0.879 {\scriptsize $\pm$0.001}$^*$ & 0.746 {\scriptsize $\pm$0.009}$^*$ & 0.662 {\scriptsize $\pm$0.006}$^*$ & 0.765$^*$\\
16. M2D~\cite{M2D2024TASLP} & 0.595 {\scriptsize $\pm$0.008}$^*$ & 0.797 {\scriptsize $\pm$0.000}$^*$ & \textbf{1.000 {\scriptsize $\pm$0.000}}$^*$ & 0.703 {\scriptsize $\pm$0.024}$^*$ & 0.905 {\scriptsize $\pm$0.001}$^*$ & 0.756 {\scriptsize $\pm$0.013}$^*$ & \textbf{0.720 {\scriptsize $\pm$0.012}}$^*$ & 0.782$^*$\\
\multicolumn{8}{l}{\textit{Ensemble SSL model (CED), large parameter model (Dasheng 1.2B), and audio-visual contrastive SSL model (OpenL3).}}  \\
17. CED~\cite{dinkel2023ced} & 0.614 {\scriptsize $\pm$0.001}$^*$ & 0.782 {\scriptsize $\pm$0.001} & 0.997 {\scriptsize $\pm$0.001}$^*$ & 0.713 {\scriptsize $\pm$0.007}$^*$ & 0.873 {\scriptsize $\pm$0.001} & 0.833 {\scriptsize $\pm$0.019}$^*$ & 0.597 {\scriptsize $\pm$0.117} & 0.773$^*$\\
18. Dasheng-1.2B~\cite{Dinkel2024Dasheng} & 0.582 {\scriptsize $\pm$0.005}$^*$ & 0.734 {\scriptsize $\pm$0.002} & 0.915 {\scriptsize $\pm$0.031}$^*$ & 0.662 {\scriptsize $\pm$0.016} & 0.772 {\scriptsize $\pm$0.002} & 0.700 {\scriptsize $\pm$0.072} & 0.660 {\scriptsize $\pm$0.021}$^*$ & 0.718\\
19. OpenL3\cite{cramer2019openl3} & 0.608 {\scriptsize $\pm$0.011}$^*$ & 0.754 {\scriptsize $\pm$0.006} & 0.978 {\scriptsize $\pm$0.007}$^*$ & 0.695 {\scriptsize $\pm$0.018}$^*$ & 0.845 {\scriptsize $\pm$0.007} & 0.751 {\scriptsize $\pm$0.027}$^*$ & 0.639 {\scriptsize $\pm$0.025}$^*$ & 0.753$^*$\\
\multicolumn{8}{l}{\textit{Respiratory audio SSL models.}}  \\
20. OPERA-CT~\cite{OPERA} & 0.578 {\scriptsize $\pm$0.001} & 0.795 {\scriptsize $\pm$0.001} & 0.855 {\scriptsize $\pm$0.012} & 0.685 {\scriptsize $\pm$0.012} & 0.874 {\scriptsize $\pm$0.000} & 0.722 {\scriptsize $\pm$0.016} & 0.625 {\scriptsize $\pm$0.038} & 0.733\\
21. OPERA-GT~\cite{OPERA} & 0.552 {\scriptsize $\pm$0.003} & 0.735 {\scriptsize $\pm$0.000} & 0.741 {\scriptsize $\pm$0.011} & 0.650 {\scriptsize $\pm$0.005} & 0.825 {\scriptsize $\pm$0.001} & 0.703 {\scriptsize $\pm$0.016} & 0.606 {\scriptsize $\pm$0.015} & 0.687\\
\midrule
\multicolumn{9}{l}{\textit{Reference from Table \ref{tab:exp:result-ablations}: M2D further pre-trained on AudioSet + Resipratory sound data. Bold results are better than the models above.}}  \\
M2D+Resp & \textbf{0.627 {\scriptsize $\pm$0.009}}$^*$ & \textbf{0.856 {\scriptsize $\pm$0.001}}$^*$ & \textbf{1.000 {\scriptsize $\pm$0.001}}$^*$ & \textbf{0.757 {\scriptsize $\pm$0.004}}$^*$ & \textbf{0.954 {\scriptsize $\pm$0.001}}$^*$ & 0.794 {\scriptsize $\pm$0.016}$^*$ & 0.714 {\scriptsize $\pm$0.007}$^*$ & \textbf{0.814}$^*$\\
\bottomrule
\addlinespace[0.03cm]
\multicolumn{9}{l}{$^*$Results better than OPERA-CT, the previous SOTA, pre-trained on respiratory sounds only.}\\
\end{tabular}
}
\vspace{-12pt}
\end{table*}

\subsection{Evaluated Models} \label{sec:exp-models}
We used audio foundation models with various backgrounds.
For models trained specifically on respiratory sounds, we used OPERA-CT/GT~\cite{OPERA}. For models trained on speech similar to respiratory sounds, we used speech self-supervised learning (SSL) models wav2vec 2.0~\cite{baevski2020wav2vec2}, HuBERT~\cite{Hsu2021HuBERT}, and WavLM~\cite{Chen2022WavLM}, which are all base models. For models trained on general audio from AudioSet, we utilized supervised learning models PANNs~\cite{kong2020panns}, AST~\cite{gong2021ast}, and HTS-AT~\cite{Chen2022HTS-AT}, as well as SSL models Audio-MAE~\cite{huang2022audiomae}, MSM-MAE~\cite{niizumi2022msm-mae}, BEATs~\cite{chen2022beats}, BYOL-A~\cite{niizumi2023byol-a}, ATST-Clip/Frame~\cite{Li2023ATST-TALSP}, and M2D~\cite{M2D2024TASLP}.
Most of the models are base-sized transformers with about 90M parameters, except for PANNs and BYOL-A.

Additionally, we examined CED~\cite{dinkel2023ced} (distills multiple Masked Autoencoders (MAE)~\cite{he2022masked}), Dasheng~\cite{Dinkel2024Dasheng} (pre-trains MAE on large datasets and parameters), OpenL3~\cite{cramer2019openl3} (a multi-modal SSL connecting video and audio), and CLAP (contrastive language-audio pre-training) models~\cite{LAION-CLAP,CLAP2022,CLAP2023}.
These models employ diverse pre-training paradigms, datasets, network architectures, and output feature aggregation techniques.

\subsection{Pre-training Dataset} \label{sec:exp-pt-dataset}
To evaluate the impact of the pre-training dataset, we utilized respiratory sound databases, a speech corpus LibriSpeech~\cite{Panayotov2015LibrispeechAA}, and a general audio dataset AudioSet~\cite{gemmeke2017audioset}. For respiratory sounds, we employed COUGHVID~\cite{COUGHVID}, HF\_Lung~\cite{HFLungV2}, and ICBHI2017~\cite{ICBHI2017}, which form a subset of the data used in OPERA. COUGHVID consists of cough sounds recorded via microphones, and we used 7054 samples with a cough detection ratio of 0.95 or higher. HF\_Lung contains lung sounds recorded via a stethoscope and another device; we used 3839 samples of stethoscope recordings. ICBHI2017 also consists of lung sounds recorded via stethoscopes, comprising 539 samples. To balance the amount of cough and lung sound data, we augmented the ICBHI2017 sample list sixfold and combined it with the lists from COUGHVID and HF\_Lung, resulting in a total of 14,127 samples used as the respiratory sound dataset.

AudioSet, widely used in general audio models, comprises various sounds, such as music, speech, and environmental sounds. It is worth mentioning that AudioSet also includes classes such as respiratory sounds, breathing, and coughing, similar to respiratory databases. We utilized 2,005,132 samples from the balanced/unbalanced train segments.

\begin{table*}[tb!]
\vspace{-10pt}
\caption{M2D ablations on OPERA benchmark.}
\label{tab:exp:result-ablations}
\vspace{-5pt}
\centering
\resizebox{0.9\textwidth}{!}{%
\begin{tabular}{lllllllll}
\toprule
\multicolumn{1}{r}{Task} & T5 & T6 & T7 & T8 & T9 & T10 & T11 & \\
Model & Covid & Gender & COPD & Smoker & Gender & Obstructive & COPD (5 cls) & Avg.\\
\midrule
M2D~\cite{M2D2024TASLP} & 0.595 {\scriptsize $\pm$0.008} & 0.797 {\scriptsize $\pm$0.000} & {1.000 {\scriptsize $\pm$0.000}}& 0.703 {\scriptsize $\pm$0.024} & 0.905 {\scriptsize $\pm$0.001} & 0.756 {\scriptsize $\pm$0.013} & 0.720 {\scriptsize $\pm$0.012} & 0.782\\

\midrule
\multicolumn{8}{l}{\textit{Summarizing output features by mean pooling instead of concatenating frequency-wise features.}}  \\
(i) Mean pooling & 0.593 {\scriptsize $\pm$0.003} & 0.728 {\scriptsize $\pm$0.002} & 0.955 {\scriptsize $\pm$0.004} & 0.642 {\scriptsize $\pm$0.015} & 0.754 {\scriptsize $\pm$0.002} & 0.728 {\scriptsize $\pm$0.030} & 0.582 {\scriptsize $\pm$0.068} & 0.712\\

\multicolumn{8}{l}{\textit{Training objective ablations.}}  \\
(ii) M2D ftAS~\cite{M2D2024TASLP} & 0.597 {\scriptsize $\pm$0.012} & 0.813 {\scriptsize $\pm$0.001} & 0.999 {\scriptsize $\pm$0.000} & 0.732 {\scriptsize $\pm$0.007} & 0.907 {\scriptsize $\pm$0.001} & 0.827 {\scriptsize $\pm$0.025} & 0.649 {\scriptsize $\pm$0.113} & 0.789\\
(iii) M2D-CLAP~\cite{niizumi24M2D-CLAP} & 0.604 {\scriptsize $\pm$0.007} & 0.809 {\scriptsize $\pm$0.000} & 0.995 {\scriptsize $\pm$0.004} & 0.737 {\scriptsize $\pm$0.032} & 0.910 {\scriptsize $\pm$0.001} & 0.752 {\scriptsize $\pm$0.039} & 0.720 {\scriptsize $\pm$0.049} & 0.790\\
(iv) M2D-S~\cite{niizumi2023m2d4speech} & 0.547 {\scriptsize $\pm$0.007} & 0.660 {\scriptsize $\pm$0.001} & 0.881 {\scriptsize $\pm$0.031} & 0.591 {\scriptsize $\pm$0.027} & 0.727 {\scriptsize $\pm$0.001} & 0.725 {\scriptsize $\pm$0.019} & 0.554 {\scriptsize $\pm$0.049} & 0.669\\

\multicolumn{8}{l}{\textit{Feature time-frame resolution ablations with patch sizes of $16\times4$ and $80\times2$.}}  \\
(v) 40ms ($16\times4$) & 0.577 {\scriptsize $\pm$0.002} & 0.787 {\scriptsize $\pm$0.001} & 1.000 {\scriptsize $\pm$0.000} & 0.706 {\scriptsize $\pm$0.007} & 0.880 {\scriptsize $\pm$0.001} & 0.738 {\scriptsize $\pm$0.018} & 0.630 {\scriptsize $\pm$0.025} & 0.760\\
(vi) 20ms ($80\times2$) & 0.582 {\scriptsize $\pm$0.005} & 0.778 {\scriptsize $\pm$0.001} & 0.956 {\scriptsize $\pm$0.006} & 0.678 {\scriptsize $\pm$0.007} & 0.882 {\scriptsize $\pm$0.003} & 0.757 {\scriptsize $\pm$0.061} & 0.663 {\scriptsize $\pm$0.135} & 0.757\\

\bottomrule\\
\end{tabular}
}
\vspace{-20pt}
\end{table*}

\section{Emperical Analysis}
We evaluated 21 audio foundation models on the OPERA benchmark (Table \ref{tab:exp:result-opera}) and conducted ablation experiments on the M2D model (Table \ref{tab:exp:result-ablations}) and on data (Table \ref{tab:exp:result-data-ablations}). We also evaluated the layer-wise performance (Figures \ref{fig:exp-layer-M2D} and \ref{fig:exp-layer-HuBERT}).

\subsection{RQ1 What are the better practices for pre-training effective respiratory audio foundation models?} \label{sec:anl-rq1}

\subsubsection{Comparing benchmark results for all models}
Table \ref{tab:exp:result-opera} shows the results for 21 audio foundation models and a reference model. We compare these models based on their pre-training settings and performance and especially focus on comparing models with the respiratory audio foundation models OPERA-CT/GT.

\vspace{0.05cm}
\noindent\textbf{Pre-training with a sufficiently large and diverse audio dataset is more advantageous than using only respiratory sound data.}
The results shows that eight out of 19 general audio models outperform the respiratory audio model OPERA-CT (The asterisks * in the table indicate a better performance than OPERA-CT).
As the eight models that outperform OPERA-CT employ a variety of pre-training backgrounds while using AudioSet, the pre-training data may have a greater impact on performance than other factors (e.g., learning methods).
Notably, while using the respiratory audio dataset should be advantageous for learning respiratory audio features, AudioSet also contains these sounds, as described in Section \ref{sec:exp-pt-dataset}.
Besides, the limited performance of OPERA-GT (which employs the SSL method MAE~\cite{he2022masked}) may indicate that the pre-training task of predicting masked parts becomes too simple with monotonous respiratory sound data, making it challenging to learn useful features.
To summarize, the observations indicate that pre-training on AudioSet's large-scale data with approximately 2M samples ($\approx$5569h) is more effective than pre-training on the respiratory sound data in OPERA with 136K samples (404.1h).

\vspace{0.05cm}
\noindent\textbf{Although speech shares the same pathway as breath sounds, speech SSLs underperform other models.}
One reason could be that these models utilize training signals obtained through clustering features of a speech signal. While they contain phonemes and other linguistic information, they are likely to have fewer non-linguistic breath features, thus making the models learn less to represent respiratory features. Similarly, BEATs, which use clustered features as training signals for pre-training on AudioSet, exhibit limited performance, suggesting that respiratory sounds are not well-represented when learned from clustered features. Furthermore, speech datasets are generally clear speech recordings, likely not to include trainable respiratory sounds, potentially resulting in the lower performance of the speech SSLs.

\vspace{0.05cm}
\noindent\textbf{SSL models learned only from audio excels across benchmark tasks.}
Models employing a wide range of learning methods have demonstrated their effectiveness. However, AST, supervised learning, CLAP$_{2023}$, contrastive learning with audio captions, and OpenL3, contrastive learning with videos, tend to perform slightly lower in gender classification tasks (T6 and T9). In contrast, models with masked prediction-based learning only from audio (ATST-Frame utilizing data augmentation, M2D enhancing MAE for training signal and prediction task, and CED distilling many MAE models) achieve consistently strong performance across tasks. Notably, ATST-Frame and M2D outperform OPERA in all tasks.

\vspace{0.05cm}
\noindent\textbf{Increasing network parameters or the amount of data used does not necessarily improve performance.}
Dasheng, pre-trained on a large-scale dataset of 1.2B parameters and 97M samples, achieves performance comparable to BYOL-A, which has 5M parameters pre-trained on a dataset with 2M samples. In the case of the CLAP model, CLAP$_{2023}$, pre-trained on 4.6M samples, shows significant performance improvement, especially in gender classification, compared to CLAP$_{2022}$, pre-trained on 128K samples. On the other hand, the top-performing models, ATST and M2D trained on the 2M-sample AudioSet dataset using about 90M parameters.

\vspace{0.05cm}
Finally, the reference result from the M2D ablation study, M2D+Resp, demonstrates a new SOTA performance on the OPERA benchmark by further pre-training M2D using respiratory sound data. The M2D+Resp shows an average result of 0.814, outperforming the former SOTA of the OPERA-CT's 0.733 with a large margin. The improvement highlights the potential for further advancements by tailoring the top-performing model specifically to respiratory sound analysis.

\begin{table*}[tb!]
\vspace{-5pt}
\caption{Data ablations of M2D on OPERA benchmark. The respiratory pre-training data contains the data from tasks with $\in$Resp.}
\label{tab:exp:result-data-ablations}
\vspace{-5pt}
\centering
\resizebox{0.9\textwidth}{!}{%
\begin{tabular}{lllllllll}
\toprule
\multicolumn{1}{r}{Task} & T5 $\in$Resp & T6 $\in$Resp & T7 $\in$Resp & T8 & T9 & T10 & T11 & \\
Pre-training: the data used$^\dagger$ & Covid & Gender & COPD & Smoker & Gender & Obstructive & COPD (5 cls) & Avg.\\
\midrule
Scratch: AS only (M2D)~\cite{M2D2024TASLP} & 0.595 {\scriptsize $\pm$0.008} & 0.797 {\scriptsize $\pm$0.000} & \textbf{1.000 {\scriptsize $\pm$0.000}}& 0.703 {\scriptsize $\pm$0.024} & 0.905 {\scriptsize $\pm$0.001} & 0.756 {\scriptsize $\pm$0.013} &  {0.720 {\scriptsize $\pm$0.012}} & 0.782\\

\midrule
Fur: Resp only & 0.612 {\scriptsize $\pm$0.004} & 0.832 {\scriptsize $\pm$0.000} & 0.990 {\scriptsize $\pm$0.008} & 0.717 {\scriptsize $\pm$0.021} & 0.922 {\scriptsize $\pm$0.001} & 0.773 {\scriptsize $\pm$0.016} & 0.635 {\scriptsize $\pm$0.022} & 0.783\\

Fur: AS+Resp 100K & 0.608 {\scriptsize $\pm$0.010} & 0.847 {\scriptsize $\pm$0.000} & 0.999 {\scriptsize $\pm$0.000} & 0.761 {\scriptsize $\pm$0.011} & 0.948 {\scriptsize $\pm$0.001} & 0.752 {\scriptsize $\pm$0.014} & 0.705 {\scriptsize $\pm$0.019} & 0.803\\
Fur: AS+Resp 200K & 0.617 {\scriptsize $\pm$0.011} & 0.848 {\scriptsize $\pm$0.001} & 0.999 {\scriptsize $\pm$0.002} & {0.758 {\scriptsize $\pm$0.004}} & 0.948 {\scriptsize $\pm$0.001} & 0.792 {\scriptsize $\pm$0.048} & 0.689 {\scriptsize $\pm$0.066} & 0.807\\
Fur: AS+Resp 300K & 0.610 {\scriptsize $\pm$0.009} & 0.851 {\scriptsize $\pm$0.001} & 0.998 {\scriptsize $\pm$0.004} & \textbf{0.768 {\scriptsize $\pm$0.004}} & 0.951 {\scriptsize $\pm$0.001} & 0.773 {\scriptsize $\pm$0.030} & \textbf{0.724 {\scriptsize $\pm$0.026}} & 0.811\\
Fur: AS+Resp 400K (M2D+Resp) & 0.627 {\scriptsize $\pm$0.009} &\textbf{0.856 {\scriptsize $\pm$0.001}}& \textbf{1.000 {\scriptsize $\pm$0.001}} & 0.757 {\scriptsize $\pm$0.004} &\textbf{0.954 {\scriptsize $\pm$0.001}}&\textbf{0.794 {\scriptsize $\pm$0.016}}& 0.714 {\scriptsize $\pm$0.007} & \textbf{0.814}\\
Fur: AS+Resp 500K & 0.632 {\scriptsize $\pm$0.006} & \textbf{0.856 {\scriptsize $\pm$0.001}} & 0.995 {\scriptsize $\pm$0.007} & {0.758 {\scriptsize $\pm$0.002}} & 0.953 {\scriptsize $\pm$0.001} & 0.760 {\scriptsize $\pm$0.021} & 0.710 {\scriptsize $\pm$0.009} & 0.809\\

Scratch: AS+Resp 400K & \textbf{0.644 {\scriptsize $\pm$0.005}} & 0.854 {\scriptsize $\pm$0.000} & 0.981 {\scriptsize $\pm$0.010} & 0.745 {\scriptsize $\pm$0.003} & 0.942 {\scriptsize $\pm$0.000} & 0.755 {\scriptsize $\pm$0.010} & 0.591 {\scriptsize $\pm$0.010} & 0.787\\
Scratch: LibriSpeech & 0.582 {\scriptsize $\pm$0.002} & 0.725 {\scriptsize $\pm$0.001} & 0.833 {\scriptsize $\pm$0.008} & 0.662 {\scriptsize $\pm$0.006} & 0.747 {\scriptsize $\pm$0.016} & 0.747 {\scriptsize $\pm$0.036} & 0.595 {\scriptsize $\pm$0.035} & 0.699\\

\bottomrule
\addlinespace[0.03cm]
\multicolumn{9}{l}{\scriptsize{$^\dagger$Scratch: pre-training from scratch; Fur: further pre-training on M2D; AS: AudioSet; Resp: the set of respiratory data and size.}}\\
\addlinespace[-0.05cm]
\multicolumn{9}{l}{\hspace{0.1cm}\scriptsize{The size of the respiratory sound data (Resp) is 14K, and the versions of 100, 200, 300, 400, and 500K were created by augmenting the data list by 7, 15, 22, 29, and 36 times, respectively,}}\\
\addlinespace[-0.05cm]
\multicolumn{9}{l}{\hspace{0.1cm}\scriptsize{to increase the proportion relative to AudioSet with 2M samples.}}\\
\end{tabular}

}
\vspace{-10pt}
\end{table*}

\subsubsection{Ablations using M2D}

We conducted ablation experiments using the top-performing model, M2D, to determine the impact on the feature aggregation, training objectives, and feature time-frame resolution, and Table \ref{tab:exp:result-ablations} shows the results.

\vspace{0.05cm}
\noindent\textbf{Preserving frequency-wise information in feature aggregation is crucial.}
Models using vision transformers divide spectrograms into $16\times16$ patches and encode each patch into features (e.g., AST, BEATs, M2D, and OPERA-GT). While most approaches aggregate the features for an input audio clip by averaging the patch features, M2D and MSM-MAE concatenate frequency-wise features for a time frame, preserving information for each frequency in the aggregated features.
The results show that (i) Mean pooling, which simply averages the M2D output features for patches, yields an average task performance of 0.712, a significant degradation from the original M2D's 0.782. In particular, tasks T6 and T9, which involve gender classification, become more challenging, suggesting that frequency components are critical for identifying gender from cough sounds. Similarly, the performance degradation in T11, which involves COPD (chronic obstructive pulmonary disease, a common lung disease) severity classification, highlights the importance of frequency components for estimating severity.
In the case of (vi) M2D with a patch size of $80\times2$, where each patch represents the entire frequency range of a single time frame, the configuration shows less degradation than (i) Mean pooling.
Similarly, ATST, which preserves the entire frequency in a patch, also shows strong performance in Table \ref{tab:exp:result-opera}.
These observations indicate that, as reported in the previous study for general audio tasks~\cite{niizumi2022composing}, preserving frequency-wise information is crucial and effective for the respiratory benchmark.

\vspace{0.05cm}
\noindent\textbf{Learning representations from the acoustic patterns of AudioSet samples is more important than learning them from its semantics or labels.}
In the training objective ablations, both (ii) M2D ftAS, an M2D further fine-tuned with AudioSet labels, and (iii) M2D-CLAP, which jointly learns from M2D and CLAP using captions, learn representations from the data distributions of labels or semantics as well as from the acoustic patterns via M2D training. However, their average performance is close to M2D's 0.782. Additionally, (iv) M2D-S, trained with speech feature clusters of LibriSpeech to learn speech patterns, shows a significant performance drop. These observations suggest that learning acoustic patterns from the AudioSet samples is more effective for the respiratory sound benchmark.

\vspace{0.05cm}
Furthermore, adjusting the patch size (v) and (vi) to narrow the temporal resolution of features does not contribute to performance improvement.

\subsection{RQ2 What data lead to learning effective features?}

We discuss the impact of the datasets on performance based on the results for M2D pre-trained on various datasets in Table \ref{tab:exp:result-data-ablations}.
We used the same setting with M2D and replaced the pre-training dataset. For further pre-training experiments, we pre-trained the AudioSet-pre-trained M2D for 50 epochs on each dataset.
Note that the respiratory sound dataset used for the pre-training contains the training data from tasks T5 to T7.

\vspace{0.05cm}
\noindent\textbf{Further pre-training on data combining AudioSet and respiratory sound datasets improves performance.}
In the experiments, we pre-trained M2D from scratch or further pre-trained it on various datasets. Among various attempts, further pre-trainings of AS+Resp 100 to 500K consistently demonstrated average performance improvement.
Further pre-training on the combination of AudioSet and respiratory sound for 400K samples (M2D+Resp) shows the best performance.

When further pre-trained using only respiratory sounds (\textit{Fur: Resp only}), the average performance becomes comparable to that of the original M2D. In this experiment, we employed the M2D-X~\cite{M2D2024TASLP} framework using AudioSet as background noise to achieve successful pre-training on a small-sized dataset, which showed the best performance in the preliminary study. The results suggest that additional training with only respiratory sounds does not further enhance the effectiveness of the representations gained from AudioSet.

Pre-training from scratch on the AS+Resp 400K dataset (\textit{Scratch: AS+Resp 400K}) improves the performance for the cough sound tasks (T5, T6, T8, and T9) while degrading the performance for other lung sound tasks. Furthermore, training solely with the speech corpus LibriSpeech in M2D also underperforms even with no feature clustering as done in M2D-S.

To summarize the observations, the most effective approach that updates the SOTA on the OPERA benchmark is to further pre-train the AudioSet pre-trained M2D on the combination of AudioSet and respiratory sound data. While this combination can refine the learned effective representations for respiratory sounds, we leave the background reason for future investigation.

\subsection{RQ3 Are the intermediate layer features effective?} \label{sec:exp-layers}

\begin{figure}[tbp]
  \centering
  \includegraphics[width=0.85\columnwidth]{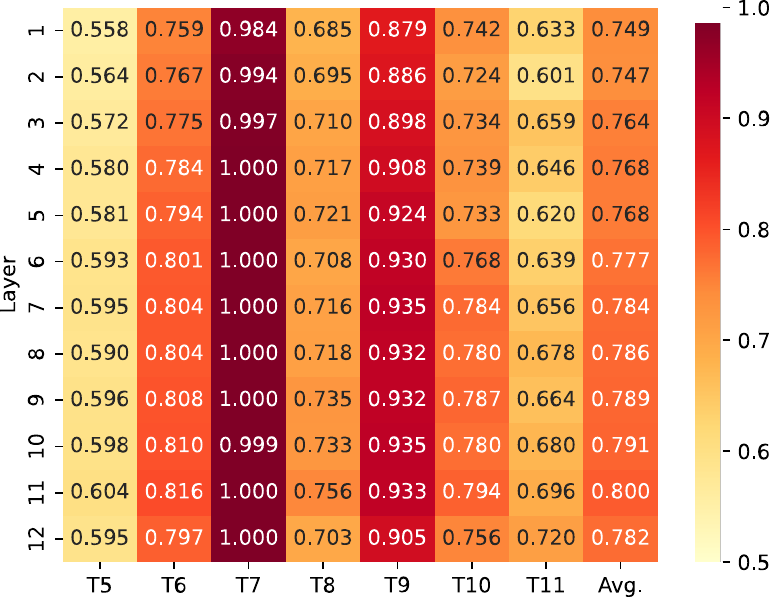}
  \vspace{-5pt}
  \caption{M2D performance by layers.}
  \label{fig:exp-layer-M2D}
  \vspace{-15pt}
\end{figure}

\begin{figure}[tbp]
  \centering
  \includegraphics[width=0.85\columnwidth]{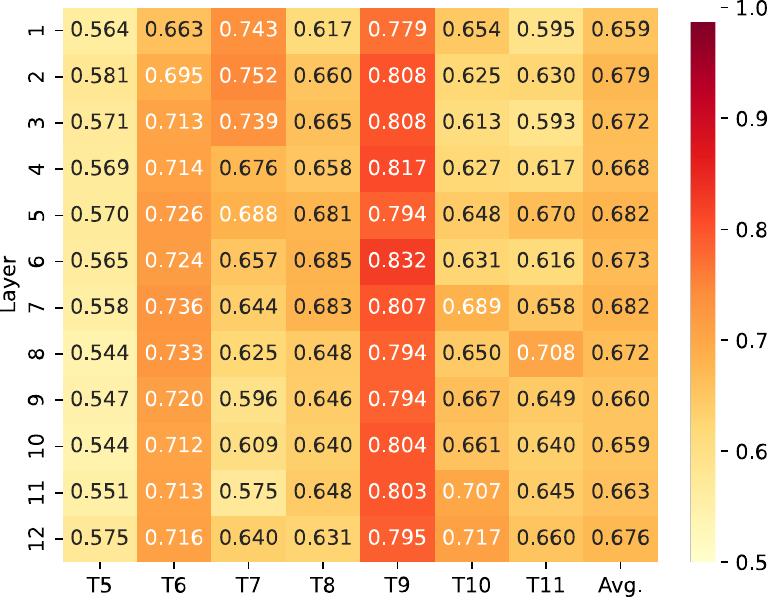}
  \vspace{-5pt}
  \caption{HuBERT performance by layers.}
  \label{fig:exp-layer-HuBERT}
  \vspace{-10pt}
\end{figure}

We evaluated each layer's output and investigated their performance on the benchmark as in the previous study \cite{niizumi2022composing}. We tested M2D from top-performing models and the best-performing speech SSL, HuBERT. We specifically added a speech SSL because layer-wise feature performance is crucial in the speech domain. Figures \ref{fig:exp-layer-M2D} and \ref{fig:exp-layer-HuBERT} shows the results.

HuBERT performs better in its earlier layers for some tasks, while the performance gain is limited.
In particular, the early layers of HuBERT perform better on task T7, while the middle layers consistently outperform the later layers on other tasks except T10. Similar to speech domain practice, utilizing features from the middle layers may lead to better performance.

In contrast, the deeper layers of M2D consistently show better performance across all tasks. This observation confirms that the typical use of the last-layer features of general audio models is also effective for respiratory sound tasks. Furthermore, the penultimate layer demonstrates the highest average performance, suggesting that leveraging this layer could be most effective for respiratory sound tasks.

\section{Conclusion}
We investigated the better practices for pre-training an effective respiratory audio foundation model by comparing numerous audio models under the unified respiratory benchmark OPERA.
Experiments provided various insights, such as the effectiveness of pre-training on general large-scale audio as well as respiratory sound datasets and the significance of preserving frequency-wise information in feature aggregation. We renewed SOTA performance on the OPERA benchmark by a large margin. Our code is available online to contribute to the progress of respiratory audio foundation models.
\vfill

\bibliographystyle{IEEEtran}
\bibliography{refs}

\end{document}